\documentclass[epj]{svjour}
\usepackage{amssymb}
\usepackage{amsmath}
\usepackage{array} 
\usepackage{epsfig}
\usepackage{graphics}
\usepackage{colortbl}
\usepackage{hhline}
\usepackage{multirow}
\begin{document}


%
\title{Total pion-proton cross-section\\ 
from the new LHCf data on leading neutrons spectra.}
\author{
R.A.~Ryutin\thanks{\emph{e-mail:} Roman.Rioutine@cern.ch}\inst{1}
}                     
%
%
\institute{{\small Institute for High Energy Physics, NRC ``Kurchatov Institute'', Protvino {\it 142 281}, Russia}}
\date{Received: date / Revised version: date}
%
\abstract{In the light of the latest data by LHCf
collaboration of the LHC
on leading neutrons spectra it is possible to
obtain total pion-proton cross-sections
in the TeV energy region. In this work
the exact extraction procedure is shown. Final
numbers for the pion-proton cross-section
are collected at several different values
of the colliding energy and compared
with some popular theoretical predictions. Errors
of results are estimated.
} 
\authorrunning{Total pion-proton cross-section from LHCf}
\titlerunning{Total pion-proton cross-section from LHCf}
\maketitle
%

\section*{Introduction}

In previous papers
we pushed forward (and discussed)
the idea of using the leading neutrons spectra at LHC to extract 
the total~\cite{ourneutrontot}, elastic~\cite{ourneutronel} and inclusive di-jet~\cite{ourneutronJJ} 
cross-sections of the $\pi^+ p$ and $\pi^+\pi^+$ scattering processes.  Actually, this 
could allow the use of the LHC as a $\pi p$ and $\pi\pi$ collider. Certainly, at LHC it 
would be difficult to measure exclusive channels 
but, instead, inclusive spectra of fast leading neutrons seem to give 
an excellent occasion to get pion cross-sections at unimaginable 
energies 1-5 TeV in the c.m.s. For further 
motivation and technical details we refer the reader to 
Refs.~\cite{ourneutrontot}-\cite{ourneutronall}.


 The process of leading neutron production has been studied at several experiments in
 photon-hadron~\cite{HERA1}-\cite{HERA5} and hadron-hadron~\cite{hadr1}-\cite{hadr5}
 colliders. 
 
  In this paper we consider process of the type $p+p\to n+X$ in the light of new data from the LHCf 
  collaboration~\cite{LHCf1}. Recently some calculations were made in~\cite{KMRn1}-\cite{workn2}. In 
  these works authors paid attention basically to the photon-proton 
  reaction, while for hadron collisions the situation was estimated to be not so 
  clear (see~\cite{workn1},\cite{workn2}).
 
  The leading neutron production is dominated by $\pi$ 
  exchange~\cite{KMRn1}-\cite{workn2} and we have a chance to extract 
  total $\pi^+ p$ cross-sections. 
 
   Since the energy becomes large, we have to take into account effects of soft rescattering which can be calculated as 
   corrections to the Born approximation. In the calculations of such absorptive effects we use Regge-eikonal 
   approach~\cite{OurApproachNew}, which is corrected by the use of new data from TOTEM~\cite{TOTEM-1}.
   
   In the first part of the paper the outlook of the method is given, while in the last section 
   this method is applied to the recent data from the LHCf collaboration~\cite{LHCf1}.
   
   The result shows that our previous proposals to use this method in CMS ZDC look rather
   realistic.

\section*{Single pion exchange and a method to extract pion-proton total cross-section}

\begin{figure}[b!]
\begin{center} 
  \includegraphics[width=0.45\textwidth]{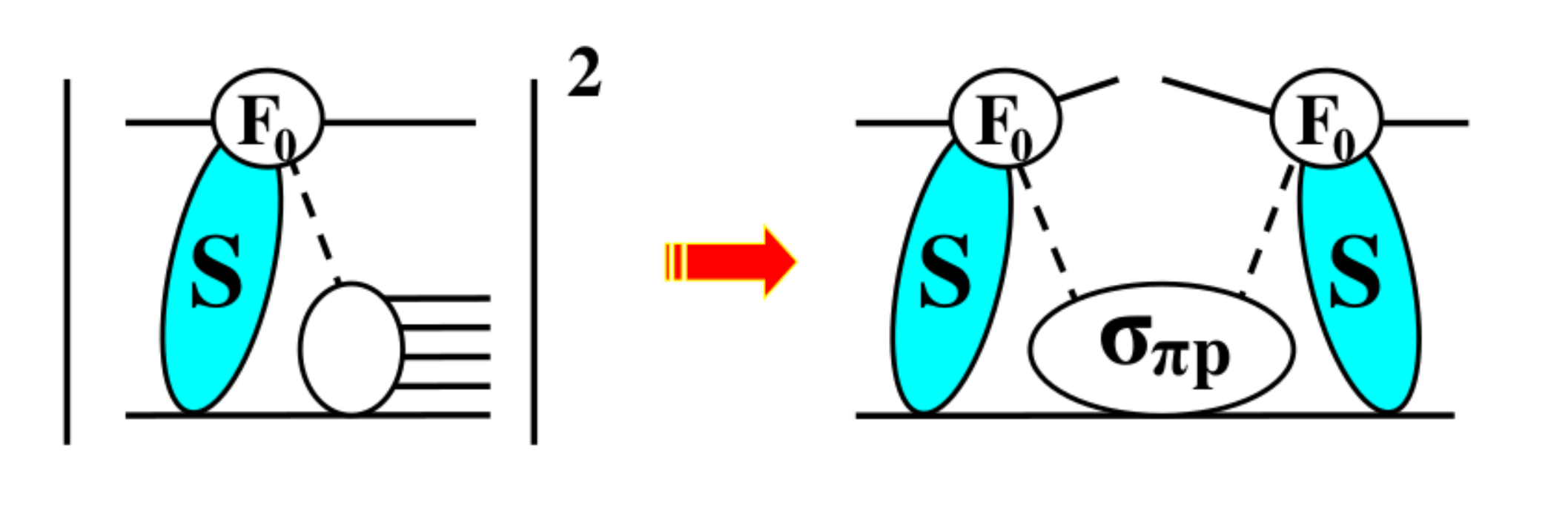}
\caption{\label{fig:diag_cs} Amplitude squared and the cross-section of the process $p+p\to n+X$ (Single pion Exchange, S$\pi$E). $S$ represents soft rescattering corrections.}
\end{center} 
\end{figure}

Details of calculations can be found in~\cite{ourneutrontot},\cite{ourneutronel}. Here we give an outlook of basic methods. As an 
approximation for $\pi$ exchanges we use the formulas shown graphically in Fig.~\ref{fig:diag_cs}. 

In the model we have to take into account absorptive corrections depicted as $S$ in Fig.~\ref{fig:diag_cs}. In our previous papers
we used the model~\cite{3Pomerons} with three pomerons for this task. In the present work we apply the Regge-eikonal model~\cite{OurApproachNew} 
with three pomerons and two odderons, since it better fits the data, including also the latest results from 
TOTEM~\cite{TOTEM-1}. Although in the region of the Single Charge Exchange (SCE) process~(\ref{eq:regionSCE}) at the LHC almost all the models
describe the data rather well, and possible theoretical errors are small. 

We consider only absorption in the initial state (elastic absorption), since other 
corrections are not so important at very low values of $t$. Arguments
in favour of this statement can be found, for example, in the Ref.~\cite{KMRn1}, where 
different types of corrections were analysed. Although, some authors~\cite{workn1},\cite{Kopeliovich1},\cite{Kopeliovich2}
argued that there is an additional suppression due to interactions of ``color octet states'' in proton 
remnants with the final neutron. But we have some doubts that the lifetime of final state fluctuations is large enough
and interaction between colorless neutron with ``color octets'' is important, at least, at low momentum transfer squared.

\begin{figure}[ht!]
	\begin{center} 
		\resizebox{0.45\textwidth}{4.5cm}{%
			\includegraphics{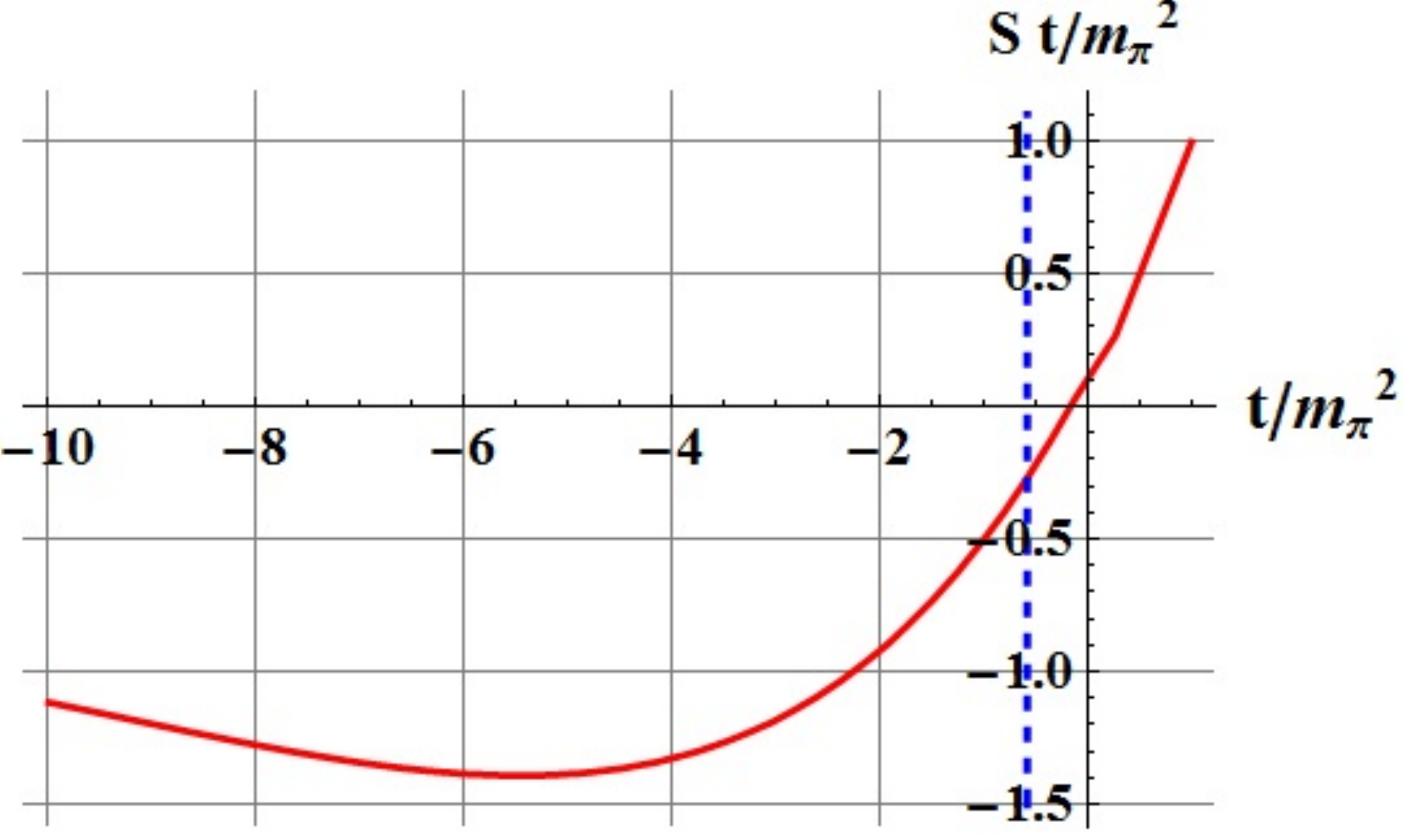}
		}
		\resizebox{0.45\textwidth}{4.5cm}{%
			\includegraphics{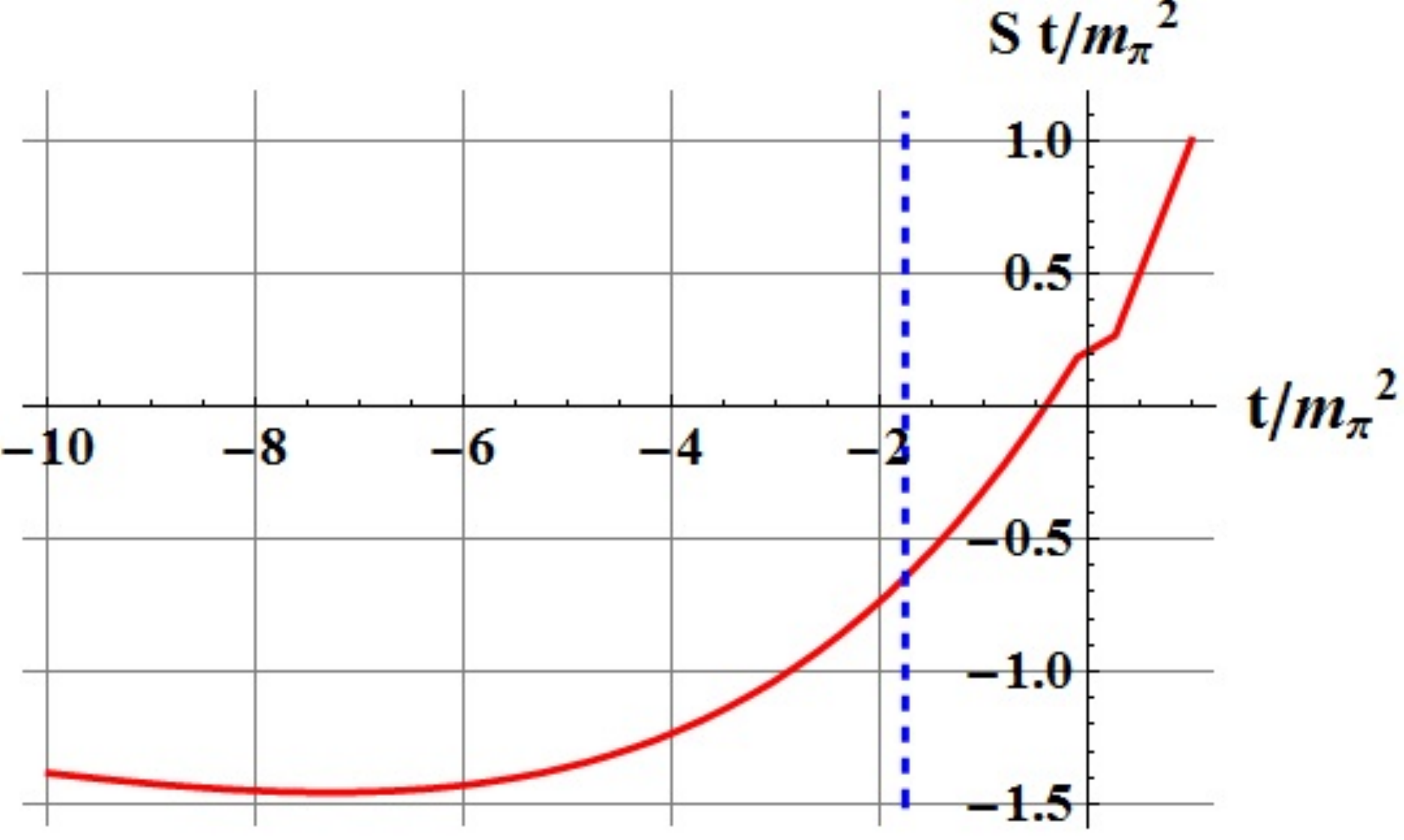}
		}
		\caption{\label{fig:St0} Function $S(\xi,t)\;t/m_{\pi}^2$ versus $t/m_{\pi}^2$ at fixed $\xi=0.107$ (upper figure) and $\xi=0.179$ (lower figure). The boundary of the physical region $t_0=-m_p^2\xi^2/(1-\xi)$ is represented by 
			vertical dashed line.}
	\end{center} 
\end{figure}

Finally we have the expression for the single pion exchange (S$\pi$E) cross-section
\begin{eqnarray}
&&\label{csSpiE}\frac{d\sigma_{{\rm S}\pi {\rm E}}}{d\xi dt}=F_0(\xi, t)S(s/s_0, \xi, t)\;\sigma_{\pi^+_{virt}p}(\xi s; \{ m_p^2, t \}) ,\\
&&\label{F0form} F_0(\xi,t)=\frac{G_{\pi^+pn}^2}{16\pi^2}\frac{-t}{(t-m_{\pi}^2)^2} {\rm e}^{2bt} \xi^{1-2\alpha_{\pi}(t)},
\end{eqnarray}
where the pion trajectory is $\alpha_{\pi}(t)=\alpha^{\prime}_{\pi}(t-m_{\pi}^2)$. The 
slope $\alpha^{\prime}_{\pi}\simeq 0.9$~GeV$^{-2}$, $\xi=1-x_L$, where $x_L$ is the fraction of 
the initial proton's longitudinal momentum carried by the 
neutron, and $G_{\pi^0pp}^2/(4\pi)=G_{\pi^+pn}^2/(8\pi)=13.75$~\cite{constG},\cite{constG2}. From recent
data~\cite{HERA2},\cite{KMRn1c16}, we expect $b\simeq 0.3\; {\rm GeV}^{-2}$. 
We are interested in the kinematical range 
\begin{equation}
\label{eq:regionSCE}
0.01\;{\rm GeV}^2<|t|<0.5\;{\rm GeV}^2,\; \xi_i<0.4,
\end{equation}
where formulae~(\ref{csSpiE}) dominate according to~\cite{KMRn1c13} and~\cite{KMRn1c14}. 

 Behaviour of $S\;t/m_{\pi}^2$ is shown in the Fig.~\ref{fig:St0}. It is clear from the figure 
that $|S|\sim 1$ at $|t|\sim m_{\pi}^2$, which is an argument for the possible
almost model-independent extraction of $\pi p$ cross-sections by the use of~(\ref{csSpiE})~\cite{ourneutronel}.
\begin{figure}[hbt!]
	\begin{center} 
		\resizebox{0.4\textwidth}{4.5cm}{%
			\includegraphics{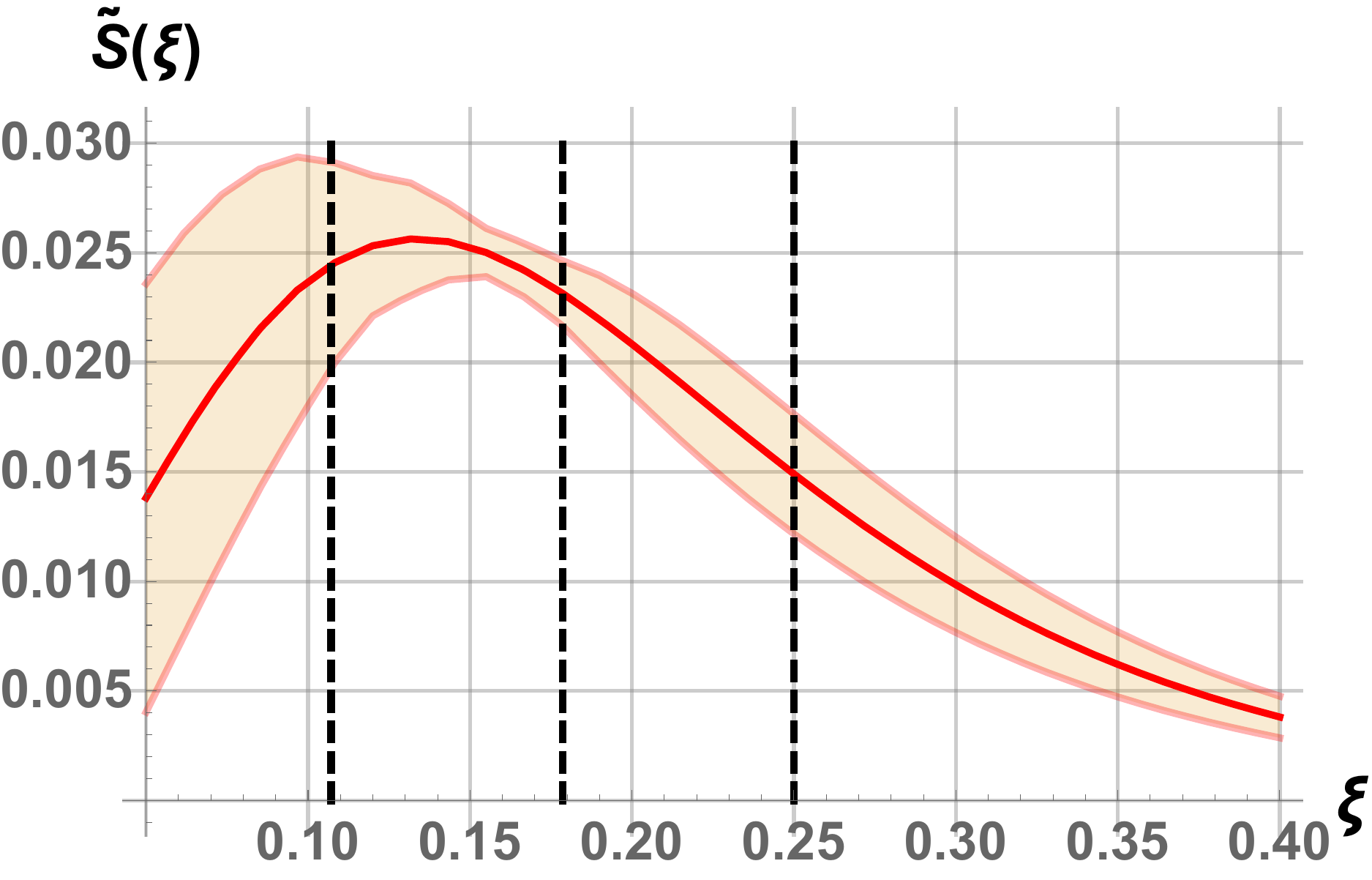}
		}
		\caption{\label{fig:survint} Rescattering corrections multiplied by formfactors for $\sqrt{s}=7$~TeV ($\tilde{S}(s,\;\xi)$) integrated in the whole $t$ regions of the LHCf data~[19]: $\eta>10.76$. Dashed vertical lines mark $\xi=0.107,0.179,0.25$, which are used to extract $\sigma_{\pi p}$ cross-sections.}
	\end{center} 
\end{figure} 

The present design of detectors does not allow exact $t$ measuremets, it gives only integrated cross-sections in some
interval $t_{\mathrm min}<t<t_{\mathrm max}$. If to 
assume a weak enough $t$-dependence of $\pi p$ cross-sections, i.e.
\begin{equation}
\label{virtpireal}
\sigma_{\pi^+_{virt}p}(s; \{ m_p^2, t \})\simeq \sigma_{\pi^+ p}(s; \{ m_p^2, m_{\pi}^2 \}), 
\end{equation}
then we could hope to extract these cross-sections (though, with big errors) by the following procedure:
\begin{eqnarray}
  \label{eq:pipextractINT}
 \tilde{S}(\xi)&\!=&\!\int\limits_{t_{min}}^{t_{max}}dt \;S(s/s_0,\xi,t)F_0(\xi,t),\\
 \label{eq:pipextractINT1} \sigma_{\pi^+ p}(\xi s)&\!=&\!\frac{\frac{d\sigma_{{\rm S}\pi {\rm E}}}{d\xi}}{\tilde{S}(s,\xi)},\; \xi\simeq\frac{M_{\pi p}^2}{s}.
  \end{eqnarray}
Function $\tilde{S}(s,\;\xi)$ is depicted in Fig.~\ref{fig:survint}. To  suppress theoretical errors of $\tilde{S}$ 
we have to use total and 
elastic $pp$ rates at $7$~TeV, since 
all the models for absorptive corrections are normalized to $pp$ cross-sections. At 
present we can estimate these errors to be less than several percents at $7$~TeV 
since we have precise data from TOTEM~\cite{TOTEM-1}.

\section*{LHCf data analysis and values of pion-proton total cross-sections in TeV domain}

\begin{figure}[bt!]
\begin{center} 
\resizebox{0.5\textwidth}{4.5cm}{%
  \includegraphics{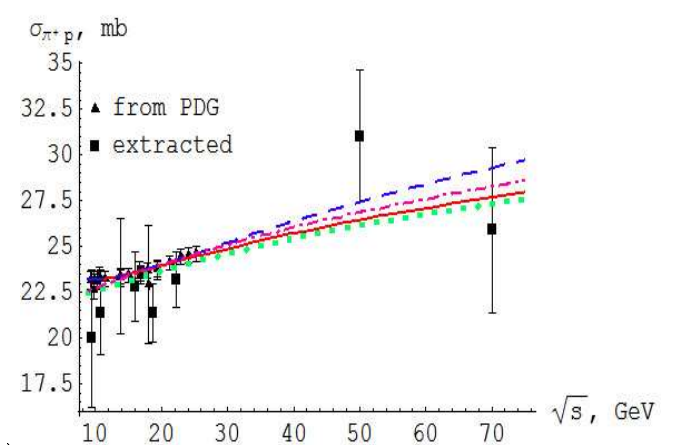}
}
\caption{\label{fig:piptotcs} Total $\pi^+p$ cross-sections versus different parametrizations:~\cite{landshofftot} (solid),\cite{COMPETE} (dashed),\cite{BSW} (dotted) and~\cite{godizov1},\cite{godizov2} (dash-dotted). Real data are taken from PDG (triangles) up to $\sqrt{s}=25\;{\rm GeV}$ and extracted values (boxes) up to $\sqrt{s}=70\;{\rm GeV}$ (see~\cite{ourneutrontot}).}
\end{center}
\end{figure}

Our method developed in~\cite{ourneutrontot}
was succesfully applied to the extraction 
of $\pi^+$ $p$ total 
cross-sections at low energies (see Fig.~\ref{fig:piptotcs}).

In this section we show results of the procedure~(\ref{eq:pipextractINT1}) applied
to the latest data on neutrons spectra by LHCf~\cite{LHCf1}.

Let us first consider the data on $d\sigma_n/dE_n$ from the table~{A.5} of~\cite{LHCf1} in three rapidity ranges:
\begin{eqnarray}
&& \eta>10.76,\nonumber\\
&& 8.99<\eta<9.22,\nonumber\\
&& 8.81<\eta<8.99.\label{eq:LHCfeta}
\end{eqnarray}
The first one corresponds to very low $t\sim m_{\pi}^2$ values, where
the flux factor $\tilde{S}$ is small, and we can use~(\ref{eq:pipextractINT1}) to extract pion-proton cross-sections. For this region we can analyze
the behaviour of functions $S\,t/m_{\pi}^2$ (Fig.~\ref{fig:St0}) and $\tilde{S}$ (Fig.~\ref{fig:survint}). In next two
regions $|t|\sim 0.1\to 0.4\;\mathrm{GeV}^2$. In principle we could calculate
$\tilde{S}$, but absorptive effects are very significant for these regions, and we should calculate them with unprecedented accuracy. This will be considered in further publications.

\begin{figure}[h!]
	\begin{center} 
		\resizebox{0.35\textwidth}{5cm}{%
			\includegraphics{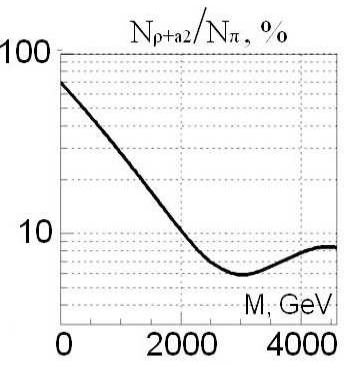}
		}
		\caption{\label{fig:BGs} Corrections (in percents) to extracted cross-sections related to additional $\rho$ and $a_2$ induced background processes in the Single Charge Exchange versus the energy of $\pi$ $p$ interaction ($M$).}
	\end{center}
\end{figure} 

\begin{table}[ht!]
	\caption{Values of the $\pi p$ total cross-sections extracted from the LHCf data~\cite{LHCf1} and also depicted in the Fig.~\ref{fig:piptotcs2}. Corresponding average $|t|$ values and $q_0$ ($q_t<q_0$) are also shown. Backgrounds from $\rho$ and $a_2$ exchanges are taken into account.}
	\label{tab:sigpiptot}       
	\begin{tabular}{llll}
		\hline\noalign{\smallskip}
		$\sqrt{s}$, TeV & $\sqrt{|t|}/m_{\pi}$ & $q_0$, GeV  & $\sigma_{\pi p}^{tot}$, mb  \\
		\noalign{\smallskip}\hline\noalign{\smallskip}
		$2.291\pm 0.382$ & $0.91\pm 0.29$  & $0.132$ & $33.15\pm 13.1$ \\
		$2.958\pm 0.296$ & $1.41\pm 0.166$  & $0.12$ & $40.22\pm 7.76$ \\
		$3.5\pm 0.25$ & $1.99\pm 0.11$ & $0.112$  & $65.43\pm 15.15$ \\
		\noalign{\smallskip}\hline
	\end{tabular}
\end{table}

\begin{figure}[hbt!]
	\begin{center} 
		\resizebox{0.45\textwidth}{4cm}{%
			\includegraphics{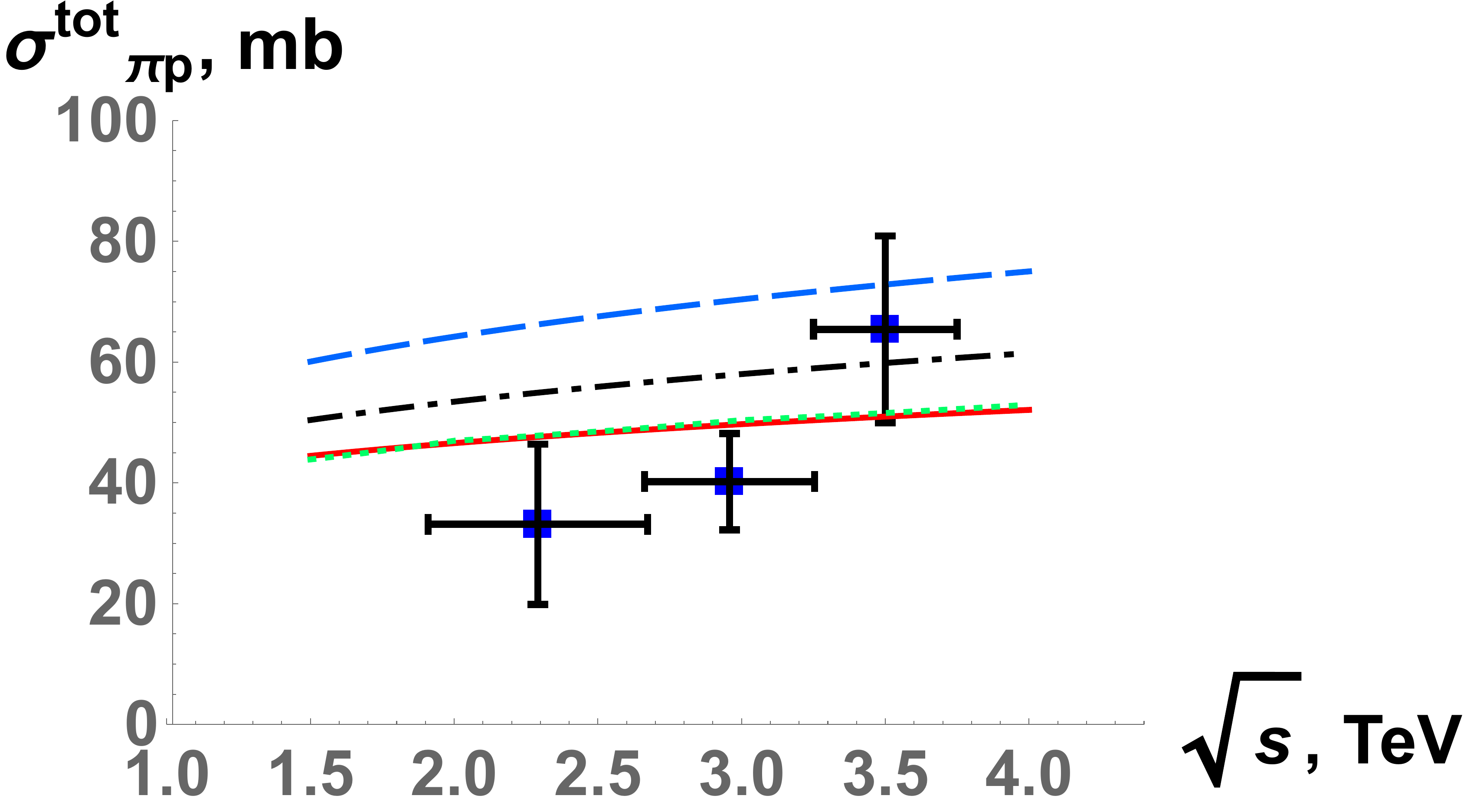}
		}
		\caption{\label{fig:piptotcs2} Extracted total $\pi^+p$ cross-sections presented in the table~\ref{tab:sigpiptot} versus different parametrizations:~\cite{landshofftot} (solid),\cite{COMPETE} (dashed),\cite{BSW} (dotted) and~\cite{godizov1},\cite{godizov2} (dash-dotted).  The interval of $t$ related to the $\eta$ region of the LHCf is $\eta>10.76$.}
	\end{center}
\end{figure} 

We also discard the data at large values of $\xi>0.25$, since the model may not work properly for large $\xi$. Finally we
use six data values from LHCf, which are reliable for our method~(\ref{eq:pipextractINT1}).

Results of calculations by the method~(\ref{eq:pipextractINT1}) are presented in the table~\ref{tab:sigpiptot} and also shown 
in the Fig.~\ref{fig:piptotcs2}. Corrections that correspond to backgrounds depicted in the Fig.~\ref{fig:BGs} are
taken into account in these results.

 Although errors of results are rather large, we can
 see following facts:
 \begin{itemize}
 	\item  Results are described well by popular models  at the 3rd point at $3.5$~TeV. But two other points are
 	at the low edge of predictions. The possible reason for the underestimation is the fact that we have to take into account some other absorptive effects (see~\cite{KMRnew} for example) that reduce $\tilde{S}$ (making extracted cross-sections higher).  
 	\item  Cross-section continues to rise with $s$.
 	\item  The pion-proton cross-section decreases
 	with $|t|$ (virtuality of the pion) increasing. Experimental errors are rather big, but preliminary calculations (which are not presented here) show the tendency. Our assumption was that this $t$ dependance is rather weak. The data confirms it rather well.
 \end{itemize}

\section*{Conclusions}

This paper was inspired by the latest LHCf data~\cite{LHCf1} on the SCE
process at 7~TeV. The analysis of these data is
the first attempt to extract
pion-proton total cross-section at TeV energies. The
observation of SCE confirms that our
expectations~\cite{ourneutrontot,ourneutronel,ourneutronJJ,ourneutronall} were realistic.
 
With the data on $p\;p$ total and elastic cross-sections at 7~TeV and higher~\cite{TOTEM-1} theoretical errors of absorptive corrections have been reduced significantly, since parameters of the model for these corrections are obtained by fitting the total and elastic cross-sections. There are
some disagreements with other authors~\cite{Kopeliovich1,Kopeliovich2}, who propose
stronger suppression factor. They considered scattering of higher Fock components
of the projectile proton, which contain a color octet 
dipole. In this case absorption occures due to pomeron exchanges
between this components and initial (final) hadron, as
depicted in the Fig.5c of Ref.~\cite{workn1}. Since
they have no calculations for single pion exchange at LHC, we can
estimate their result from calculations for double 
pion exchange in~\cite{Kopeliovich2}. They use the flux factor,
which is equal to our function $\tilde{S}$ with 
$|t_{min}|\simeq m_p^2\xi^2/(1-\xi)$ and $|t_{max}|=\infty$
in~(\ref{eq:pipextractINT}). In their case $\tilde{S}$ is approximately
15\% smaller than our result. So we can suppose that 
extracted values of the pion-proton cross-sections 
will be about 15\% higher than in the 
table~\ref{tab:sigpiptot}. 

Since calculation of absorptive corrections is the critical point, we will
discuss this question in details further, especially 
in processes like $\gamma^*\;p\to X\;n$ or $\gamma^*\;p\to\rho\;\pi\;n$,
where we have experimental data.

Unfortunately, experimental errors of the LHCf are 
huge. Nevertheless, we can try our method to extract the
pion-proton total cross-section in the TeV energy region and
make preliminary conclusions on its behaviour
at different values of
$t$.

If measurements are done more accurately then we will have additional, more rich, data in the high energy region to check predictions of different models for strong interactions, quark counting rules, "asymptopia" hypothesis 
and so on. 

\section*{Acknowledgements}

I am grateful to Vladimir Petrov for useful discussions. Thanks also to Mikhail Ryskin for detailed comments to the article. The new paper by KMR on this subject~\cite{KMRnew} was also useful.

\end{document}